\pdfoutput=1
\documentclass[geosciences,review,accept,oneauthor,pdftex,10pt,a4paper]{mdpi} 

\firstpage{1} 
\pubvolume{8}
\issuenum{10}
\articlenumber{365}
\pubyear{2018}
\copyrightyear{2018}
\history{Received: 3 July 2018; Accepted: 26 September 2018; Published: 29 September 2018}
\usepackage{amssymb}
\usepackage{hyperref}
 \theoremstyle{mdpi}
 \newcounter{thm}
 \newcounter{ex}
 \newcounter{re}
\Title{Microlensing Searches for Exoplanets}

\Author{Yiannis Tsapras}

\AuthorNames{Yiannis Tsapras}

\address [1]{%
Zentrum f{\"u}r Astronomie der Universit{\"a}t Heidelberg, Astronomisches Rechen-Institut, M{\"o}nchhofstr. 12--14, 69120 Heidelberg, Germany; ytsapras@ari.uni-heidelberg.de\\
}

\corres{Correspondence: ytsapras@ari.uni-heidelberg.de; Tel.: (+49)-(0-)6221-54-1819}

\abstract{Gravitational microlensing finds planets through their gravitational influence on the light coming from a more distant background star. The presence of the planet is then inferred from the tell-tale brightness variations of the background star during the lensing event, even if no light is detectable from the planet or the host foreground star. This review covers fundamental theoretical concepts in microlensing, addresses how observations are performed in practice, the~challenges of obtaining accurate measurements, and explains how planets reveal themselves in the data. It~concludes with a presentation of the most important findings to-date, a description of the method's strengths and weaknesses, and a discussion of the future prospects of microlensing.}


\begin{document}

%%%%%%%%%%%%%%%%%%%%%%%%%%%%%%%%%%%%%%%%%%
%% Sections that are not mandatory are listed as such. The section titles given are for Articles. Review papers and other article types have a more flexible structure. 

%% Only for the journal Gels: Please place the Experimental Section after the Conclusions

%%%%%%%%%%%%%%%%%%%%%%%%%%%%%%%%%%%%%%%%%%
%\setcounter{section}{-1} %% Remove this when starting to work on the template.

\section{Introduction}
\vspace{-6pt}

\subsection{Discovering Exoplanets}
Planets begin forming in the gaseous disks surrounding young stars, but the exact physical processes that drive their formation and evolution are still not fully understood \cite{2014PNAS..11112616F}. Theory predicts that these disks should last for a few million years and that planetary embryos can migrate while they are still embedded in them \cite{2012ARAA..50..211K}. Some are eventually engulfed by their host stars. Others survive and grow by accreting gas and dust from the surrounding disk but end up far from where they first started forming. A question may then be posed: {\it How are these planets distributed around their host stars and are there similarities with our own system or is the Solar system in some ways unique?}

The presence of liquid water on a planetary surface is no guarantee for the existence of life, although it was a necessary ingredient for the development of life on Earth. Liquid water can only exist at a certain range of distances from the host star, the so-called {\it habitable zone}, but little is known about the fraction of planets residing within the habitable zone of their host stars and the stability of their orbits; nor is it yet possible to make definite statements about the physical characteristics of even the closest exoplanets \cite{2018NatAs...2..214D,2016Natur.536..437A}. The situation becomes even more challenging when looking for planets beyond the snow line, which is the boundary distance from a star beyond which the ambient temperature drops below $\sim$160 K and water turns to ice \citep{2006ApJ...640.1115L,2011Icar..212..416M}. The exact location of the snow line in a protoplanetary disk depends on the mass accretion rate but it roughly corresponds to distances of a few AU away from the host. However, the sensitivity of the most successful planet-detection methods to-date drops rapidly for orbits $\gtrsim$1 AU, making these planets particularly hard to find.

Discovering planets and studying their distribution is the first goal of the exoplanet community, from both the observational and theoretical perspectives, in order to better understand their diversity and the physical processes that drive their formation and evolution. To this effect, a number of different techniques have been developed and employed over the past two decades. They are as follows:
\begin{itemize}[leftmargin=*,labelsep=5.5mm]
\item{{\it Radial velocity} detects planets by measuring the periodic shifting of spectral lines on the stellar spectrum that is induced by the orbital motion of the planet around the host star. Discoveries from radial velocity surveys over the past 20 years provide the first evidence that the incidence of giant planets increases with increasing stellar mass, at least for planets with short orbital periods \cite{2008PASP..120..531C}. Although the radial velocity method has been remarkably successful in identifying planets near their hosts, discoveries of giant planets beyond 1 AU for low-mass stars have been comparatively few \cite{2014ApJ...781...28M} because detecting planets at larger orbital distances requires mission lifetimes that span several years to decades.}
\item{{\it Transit} surveys, from the ground and from space, identify exoplanets with near edge-on orbits that pass in front of their host stars causing them to appear dimmer for the duration of the event. The most successful of these surveys to date has been the Kepler space mission \cite{2013ApJS..204...24B}, having found hundreds of large and small planets out to $\sim$1 AU from their hosts and thousands of candidates. Although its primary mission was suspended in 2013 due to the failure of two reaction wheels, without which the telescope could not point accurately, it provided strong evidence that terrestrial planets are far more numerous than gas giants for periods less than 85 days \cite{2013ApJ...766...81F}.}
\item{{\it Astrometry} involves very precise measurements of a star's position in the sky over a long period of time. If the star has planets orbiting it, then minute periodic shifts in its measured position could be detectable. The difficulty lies in extracting such highly accurate measurements, because the expected shifts are minuscule. This is the reason that no astrometric planet candidates have been confirmed to date. The Gaia space mission may change all that. It launched on 19 December 2013, and it is estimated it will discover thousands of planets in its expected 5-to-10-year lifetime~\cite{2014ApJ...797...14P}. It~is worth noting that the strength of the astrometric signal is inversely proportional to the distance of the planet and its host star from the Earth.}
\item{{\it Pulsar timing} provided the first successful exoplanet detections around Pulsars in 1992 \cite{1992Natur.355..145W}. Pulsars are rapidly rotating neutron stars; they are the superdense stellar remnants of supernova explosions in the distant past. The beam of electromagnetic radiation they emit as they rotate is detected on the Earth as it sweeps by, and is recorded as a highly regular and ultra-precise pulsation signal. Planets orbiting the pulsar reveal themselves through irregularities in the arrival time of the pulses.}
\item{{\it Direct imaging} is the only method aiming to visually isolate an exoplanet from its host star. Through high-contrast imaging using adaptive optic systems on large telescopes, massive exoplanets orbiting at the outer reaches of their systems have been discovered \cite{2015Sci...350...64M}. Only those planets that lie very far from their host star can be found with this method, because a star will outshine any nearby planets by a billion times in the optical part of the spectrum.}
\item{{\it Microlensing}, in many ways the odd one out, is the topic of this review. It discovers planets by measuring characteristic variations in the brightness of a background star generated by the gravity of intervening objects along the line of sight.}
\end{itemize}

\begin{figure}[H]
\centering
\includegraphics[width=12cm]{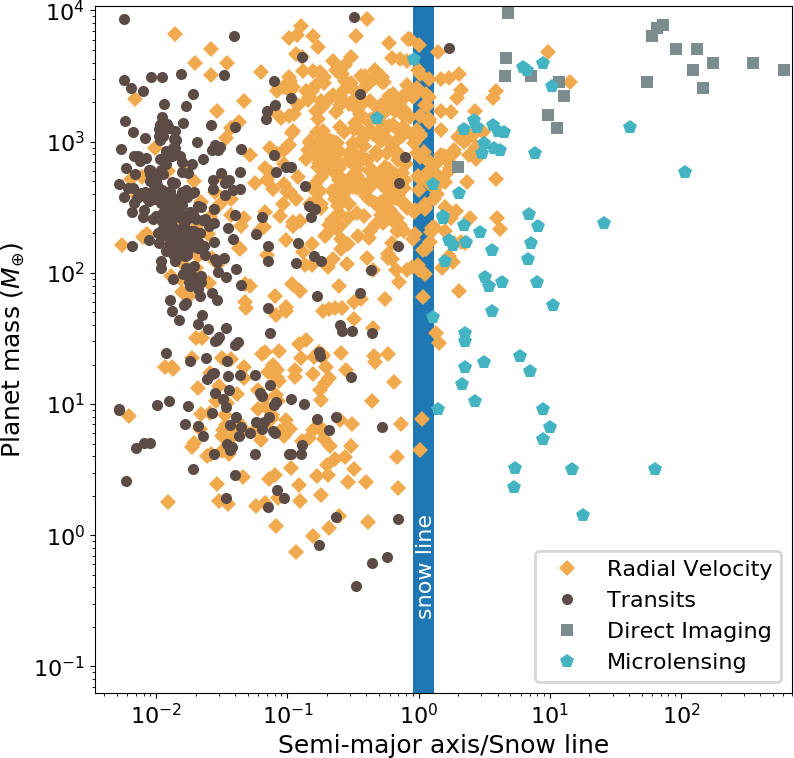}
\caption{Planet-finding techniques are complementary. This ``orbit size-vs-planet mass'' diagram shows all reported exoplanets to-date. The semi-major axis has been scaled to the approximate location of the snow line of the planet-hosting star (assuming $a_{snow} \approx 2.85 M_*^{3/2}$AU). Transits and radial velocity are exceptionally good at finding ``hot'' planets close to their host stars, whereas microlensing and direct imaging are more efficient in discovering ``cold'' planets. \it{[Figure based on data from the NASA Exoplanet Archive.]}}
\label{fig:sensitivities}
\end{figure}

These are all complementary techniques that explore different regimes of the planet mass versus orbit size distribution (see Figure \ref{fig:sensitivities}). In order to better appreciate the physical processes that form these exoplanets, theoretical work is informed and validated through new discoveries over the full range of planet mass and orbital separation. Transit missions and radial velocity surveys have already established that the abundance of hot planets increases dramatically towards lower mass objects and larger orbital separations for FGK-host systems \cite{2011arXiv1109.2497M}. There is also some evidence from microlensing that the planet frequency continues to rise towards smaller planet masses at even larger orbital radii~\cite{2012Natur.481..167C}: $17^{+6}_{-9}$\% of stars host a cool Jovian-type (0.3--10 $M_{\mathrm{Jupiter}}$) planet, $52^{+22}_{-29}$\% have cool Neptunes (10--20 $M_{\mathrm{Earth}}$) and as many as $62^{+35}_{-37}$\% have cool super-Earth (5--10 $M_{\mathrm{Earth}}$) companions. This hints that there is a still untapped large population of low-mass, cool planets, as predicted by planet population synthesis models \cite{2013ApJ...775...42I}, but the evidence remains inconclusive since very few planets have been discovered beyond the snow line. Furthermore, a recent study by \citet{2016ApJ...833..145S} found that the distribution of mass ratios and separations of their microlensing planet sample was better fit with a broken power-law model, implying that cold Neptunes could well be the most common type of planets beyond the snow line. The question is far from settled.
These are the planets microlensing is uniquely sensitive to, and finding them is critical in understanding the physical processes that drive planet formation. 

\subsection{What is Microlensing?}
\label{sec:whatis}
According to Einstein's Theory of General Relativity, gravity may be thought of as a curvature in space-time. Any massive stellar object will therefore deflect the path of light-rays passing close to it, as famously demonstrated by Eddington's observations during the total Solar eclipse of \mbox{29 May 1919 \cite{1911AnP...340..898E,1915SPAW...47..831E,1920RSPTA.220..291D}}. This phenomenon is called {\it gravitational lensing} and it involves a foreground stellar mass object bending light due to its gravity and a background stellar mass object which acts as the source of light. Any sufficiently massive object can act as a gravitational lens, from single stars to entire galaxies, and any object emitting radiation can act as the source. For reasons of convenience, the~object doing the lensing is commonly simply referred to as the {\it lens}, while the background object is the {\it source}. Gravitational lensing differs from conventional optical lensing in that there is not a single point of focus; the bending becomes more severe the closer the light-ray passes to the lens. As a consequence, multiple distorted images of the source appear around the lens. The number of images produced depends on the mass distribution doing the lensing. Under the assumption of point lenses, a~single lens will always produce two images, whereas a binary lens will generate three or five images, depending on the position of the source relative to the lens on the plane of the sky. In the special case where lens and source are in perfect alignment as viewed from the observer's perspective, the multiple images all merge to form a bright ring around the lens, the so-called `Einstein ring'. The radius of this ring, as we shall discuss in Section \ref{theory}, depends on the relative distances between observer, lens and source, as well as the mass of the lens itself.

\textls[-15]{{\it Gravitational microlensing}, or simply {\it microlensing}, is a sub-category of gravitational lensing in which both lens and source are stars and where the angular distances between the images generated by the lensing effect are of the order of milli-arcseconds. This means that no telescope in operation today can individually resolve the images; only variations in the brightness of the source are observed. Prompted by a conversation he had with Rudi Mandl, a Czech engineer and amateur scientist, Einstein addressed this phenomenon in a short article he published in 1936 \cite{1936Sci....84..506E}. In that article he worked out the basic equations and concluded that ``{\it there is no great chance of observing this phenomenon}''. In fact, \citet{1997Sci...275..184R} report that Einstein had already derived the relevant mathematical relations in his research notebooks dating back to 1912 (3 years before completing his general theory of relativity), but~had never sought to publish them thinking them of little scientific value. This was not unreasonable given the technology available at the time. Modern estimates of the optical depth to gravitational microlensing, i.e., the~probability that any given star is microlensed at any particular time, are of order $10^{-6}$ in the direction of the Galactic bulge~\cite{1994ApJ...430L.101K,1994AcA....44..165U,1998ApJ...509..177P,2013ApJ...778..150S}. Finding microlensing events is like looking for the proverbial needle in the haystack. Yet~with today's wide-field CCD cameras, astronomical surveys monitor regularly about a billion stars and discover about 2000 microlensing events annually \cite{2016JKAS...49...37K,2015AcA....65....1U}. It is worth noting that the number of reported events and the optical depth are not independent quantities, as we shall see in Section~\ref{sec:opdepth}; they~are related by the event duration. Fewer than 10 of these events every year will display the unmistakable signs of planetary companions to the lens star, a situation unlikely to change much until the launch of the WFIRST space mission in the mid-2020s \cite{2016MNRAS.455.3656P,2016AA...595A..53B}.}

\section{Microlensing Basics}
\textls[-15]{This section covers the theoretical background of the microlensing method, without going into excessive detail. Readers interested in the fundamental equations, the geometry and phenomenology of microlensing will find here enough information to gain a good understanding of the particulars of this technique. For those that want to delve deeper, the book {\it Gravitational Lenses} by~\mbox{\citet{1992grle.book.....S}} provides detailed derivations and investigates the underlying physics. Past~reviews by~\mbox{\citet{2010GReGr..42.2075D}},~\mbox{\citet{2012RAA....12..947M}}, \citet{2012ARA&A..50..411G} and the book {\it Gravitational Lensing in Astronomy}~by \citet{1998LRR.....1...12W} are also excellent sources of information. Students and researchers wanting to learn more about microlensing can access educational resources and training exercises on-line at \url{http://microlensing-source.org/}. Casual readers not interested in the mathematics may skip this section without loss of continuity.}

Note that the thin-screen and small-angle approximations are used throughout this section, though not explicitly mentioned. The thin-screen approximation assumes that the deflection occurs instantaneously during the crossing of the lens plane (at $D_L$). This is valid since the distances to the source and the lens are much greater than the range within which the actual deflection takes place. The small-angle approximation is also valid because all the angles we are dealing with are very small for all cases of interest.
\label{theory}

\subsection{Single Lens}
Gravitational microlensing will occur any time two stars are sufficiently well aligned so that the light from the background `source' star, at distance $D_S$ from the observer, will bend under the influence of the gravitational field of the intervening `lens' star, at distance $D_L$, and make the source appear temporarily brighter. As previously mentioned, microlensing events are extremely rare occurrences, so~most modern microlensing surveys point their telescopes toward the very dense stellar fields of the Galactic bulge in order to maximize their chances of finding them \citep{ogle,moa,kmtnet}. Together, low-mass K and M-dwarf stars make up about 90\% of all stars in the Galaxy. Their relative abundance is at least partly reflected in the type of lensing stars the surveys detect; indeed, most lensing stars turn out to be faint K and M-dwarfs, rather than the hotter G-type stars, like our Sun.

The basic concepts of microlensing are easier to introduce by considering a simple scenario: a~point lens in the foreground bending the light of a point background source. This is commonly referred to as ``point source-point lens'' (PSPL) lensing \cite{1936Sci....84..506E,1986ApJ...304....1P,2009MNRAS.396.2087H,2018A&A...609A..55H}. It is a sufficiently good approximation for the illustrative cases we will be considering here, but let us note now that for certain geometric arrangements the finite extent of the source star becomes important. Such considerations are covered in Section \ref{sec:second_order1}. 

\subsubsection{The Deflection Angle}
\begin{figure}[H]
\centering
\includegraphics[width=14cm]{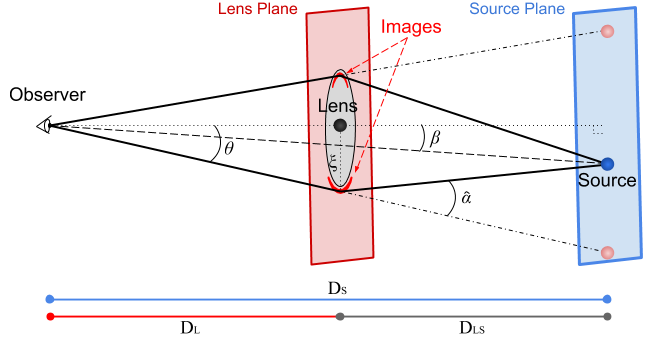}
\caption{The geometry of a single lens microlensing event. Light rays originating from a source star are deflected near the lens. As a result, the observer does not perceive the light of the source itself, but~instead measures the combined contribution of two distorted images of the source. The distances from the observer to the lens and the source are $D_L$ and $D_S$, respectively. The distance between lens and source is $D_{LS}$. $\beta$ is the angular position of the unlensed source, $\alpha$ denotes the deflection angle and $\theta$ is the angular position of one of the images. The red and blue rectangles represent the parallel lens and source planes at $D_L$ and $D_S$, respectively, which are perpendicular to the observer-lens axis (dotted).}
\label{fig:single_lens_side}
\end{figure}

\textls[-15]{The basic geometry of microlensing is illustrated in Figure~\ref{fig:single_lens_side}: Were an observer able to visually resolve arbitrarily small angular separations in the sky, they would not be able to see the source star itself, but only two distorted images generated by the lensing effect. In reality, the images are so closely packed together that they appear as a single object on the detector. According to general relativity, a light ray passing close to a lens of mass $M_L$ at distance $\xi$, will be deflected by the ``Einstein angle'' \citep{1915SPAW...47..831E,1992grle.book.....S}}
\begin{equation}
\label{eqn:def_ang}
\hat{\alpha} =\frac{4GM_L}{c^2\xi},
\end{equation}
where $c$ is the speed of light in vacuum and $G$ the gravitational constant.

\subsubsection{The Lens Equation}
Let us consider Figure~\ref{fig:single_lens_side} again. A bright source at distance $D_S$ is being observed from the Earth. Between the observer and the source lies a lens at distance $D_L$ from the observer. The distance between lens and source is $D_{LS}$. The angle between the true position of the source (in the absence of lensing) is $\beta$, whereas the angle between the observer-lens optical axis and the image considered is $\theta$. These quantities are related by
\begin{equation}
\beta = \theta - \hat{\alpha}\frac{D_{LS}}{D_S}.
\end{equation}

This is commonly referred to as the {\it lens equation}. Given an image position $\theta$, we can use the lens equation to evaluate the real position of the source, $\beta$.

\subsubsection{The Einstein Radius}
Under our assumption of a point-mass lens, we can substitute in $\hat{\alpha}$ and write the lens equation as
\begin{equation}
\label{eqn:beta2}
\beta = \theta - \frac{D_{LS}}{D_SD_L}\frac{4GM_L}{c^2\theta},
\end{equation}
where were have used the substitution $\xi = \theta D_L$. 

Since the system is rotationally symmetric, a source located exactly on the optical axis ($\beta = 0$) will be imaged as a ring: the Einstein ring. By setting $\beta = 0$ in Equation (\ref{eqn:beta2}) and solving for $\theta$, we obtain the expression for the angular Einstein radius:
\begin{equation}
\label{eqn:einstein_radius}
\theta_E = \sqrt{\frac{D_{LS}}{D_SD_L}\frac{4GM_L}{c^2}}.
\end{equation}

The angular Einstein ring radius provides a natural scale for describing the lensing geometry. The~angular separation between the images of the source generated by the lensing effect is of the order of 2$\theta_E$. The closer the angular distance of the source to the observer-lens optical axis, the stronger the magnification. In order to derive an average estimate for the size of $\theta_E$, we can assume some typical values \citep{1986ApJ...304....1P,1991ApJ...371L..63P}. Consider an M-dwarf lens star ($M_L = 0.3 M_{\odot}$) at a distance of $D_L=6.5$ kpc from the Earth, and a source star at a distance $D_S=8.5$kpc. The corresponding size of the angular Einstein ring radius is then approximately
\begin{equation}
\label{eqn:theta}
\theta_E \approx 0.902\mathrm{mas}\left(\frac{M_L}{M_{\odot}}\right)^{1/2}\left(\frac{10\mathrm{kpc}}{D_L}\right)^{1/2}\left(1-\frac{D_L}{D_S}\right)^{1/2} \sim 300 \mu\mathrm{as}.
\end{equation}

This corresponds to a physical distance on the lens plane of 
\begin{equation}
r_E = \theta_E D_L \sim 3\mathrm{AU}\left(\frac{M_L}{M_{\odot}}\right)^{1/2} \sim 1.6\mathrm{AU}.
\end{equation}

\subsubsection{Image Positions}
We can now simplify Equation (\ref{eqn:beta2}) by substituting in $\theta_E$, and the lens equation becomes \mbox{$\beta = \theta - \theta_E^2/\theta$}. Multiplying both sides with $\theta$ yields a quadratic: $\theta^2 - \beta\theta - \theta_E^2 = 0$ with solutions
\begin{equation}
\label{eqn:impos}
\theta_{\pm}=\frac{1}{2}\left(\beta \pm\sqrt{\beta^2+4\theta_E^2}\right).
\end{equation}

\begin{figure}[H]
\centering
\includegraphics[width=14cm]{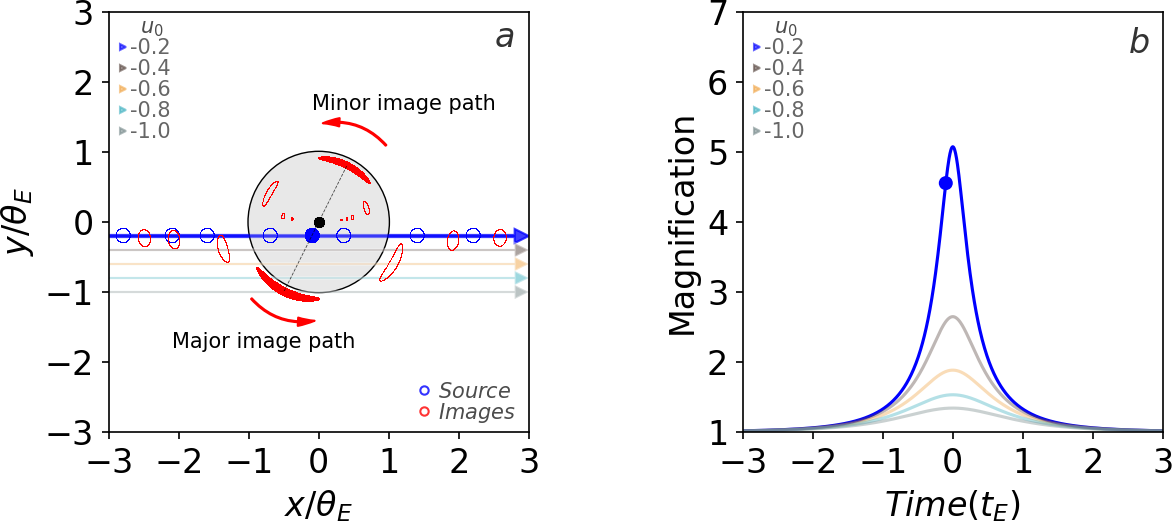}
\caption{(\textbf{a}) The face-on geometry of a single lens microlensing event. The lens is the black dot at the center. The Einstein ring is indicated by the edge of the larger gray circle. Source trajectories for five different impact parameter values, $u_0$, are shown. For the trajectory passing closest to the lens (blue line), 8 instantaneous positions for the source (unfilled blue circles) and their corresponding images (red contours) are displayed. For the source position highlighted (filled blue circle), the two images generated are shown as filled red contours. The image centers always fall in a straight line that passes through the position of the lens, as indicated by the dotted line. The minor images move along inside the Einstein ring as the event progresses, while the major images are always outside. Smaller lens-source separations on the plane of the sky correspond to greater total magnifications. The $x$ and $y$-axes mark distances on the lens plane in units of $\theta_E$; (\textbf{b}) Total magnification as a function of time for the trajectories shown in panel \textbf{a}. The filled blue circle marks the instantaneous magnification when the source is at the corresponding position (filled blue circle) in panel \textbf{a}.}
\label{fig:single_lens}
\end{figure}

These solutions correspond to the angular positions of the images on the sky. They appear close to the Einstein ring on opposite sides of the lens; the minor image always inside the Einstein ring and the major image outside. The angular separation between them is $\Delta\theta = \theta_+ - \theta_- = \sqrt{4\theta_E^2 + \beta^2} \geq 2\theta_E$. As the apparent lens-source separation increases with time, the minor image gets smaller and moves ever closer to the position of the lens, until it disappears. On the other hand, the major image gradually approaches the true position of the source until it coincides with it in terms of position and brightness (see Figure \ref{fig:single_lens}).

\subsubsection{Magnification}
\label{sec:pspl_mag}
Gravitational light deflection does not involve emission or absorption, so the surface brightness of either image is identical to that of the original source: surface brightness is conserved. The total flux received from an image is just the product of its surface brightness with the area it covers on the sky. Since the former remains unchanged during microlensing, the brightness change (magnification) of an image relative to the unlensed source, is simply given by the ratio of their areas (see Figures \ref{fig:single_lens_side} and \ref{fig:single_lens}):
\begin{equation}
A_{\pm} = \left| \frac{\theta_{\pm}}{\beta} \frac{d\theta_{\pm}}{d\beta} \right|.
\end{equation}

Substituting $\theta_{\pm}$ from Equation (\ref{eqn:impos}), we obtain:
\begin{equation}
A_{+} = \frac{\theta_{+}}{2\beta}\left(\frac{\beta}{\sqrt{\beta^2+4\theta_E^2}}+1\right),\quad
A_{-} = \frac{\theta_{-}}{2\beta}\left(\frac{\beta}{\sqrt{\beta^2+4\theta_E^2}}-1\right).\end{equation}

By normalizing all angles by $\theta_E$ and defining $u=\beta/\theta_E$ for the (normalized) source position, $y_{\pm} = \theta_{\pm}/\theta_E$ for the (normalized) image positions we can rewrite Equation (\ref{eqn:impos}) as
\begin{equation}
y_{\pm}=\frac{u\pm\sqrt{u^2+4}}{2}.
\end{equation}

The magnification of each image is then given by
\begin{equation}
A_{+} = \frac{u^2+2}{2u\sqrt{u^2+4}} + \frac{1}{2},\quad 
A_{-} = \frac{u^2+2}{2u\sqrt{u^2+4}} - \frac{1}{2}.
\end{equation}

Note here that the first term on the right-hand side of the equation above is always $>1/2$ and the minor `-' image appears inverted, as shown in Figure \ref{fig:single_lens}. The total magnification is given by the sum of the individual magnifications:
\begin{equation}
\label{eqn:magn}
A = A_+ + A_- = \frac{u^2+2}{u\sqrt{u^2+4}}.
\end{equation}

In the special case when the source lies exactly at the Einstein radius, we have $\beta=\theta_E$, $u=1$, and~the source is magnified by $A=3/\sqrt{5}\approx 1.34$. In general, the total flux $F(t)$ recorded by the detector is not proportional to the magnification because it contains contributions from other sources of light. These are discussed in Section \ref{sec:blend}.

The magnification is a function of time since lens and source are moving relative to each other with a relative proper motion $\mu_{LS}=v_{\perp}/D_L$, where $v_{\perp}$ is the relative transverse velocity of the lens with respect to the source. This gives rise to the characteristic shape of the event light curve seen in Figure \ref{fig:single_lens}b, i.e., of how the brightness of an event changes with time. This is sometimes referred to as the Paczy\'{n}ski curve \citep{1986ApJ...304....1P}. In the simplest case, the lens-source trajectory can be assumed to be rectilinear, and we may write
\begin{equation}
u(t) = \sqrt{u_0^2 + \left(\frac{t-t_0}{t_E}\right)^2},
\end{equation}
\textls[-20]{where $u_0$ is the {\it minimum impact parameter} of the event, when the apparent separation between lens and source is shortest and the magnification is greatest (denoted by $A_0$), $t_0$ is the time when $u = u_0$ and corresponds to the time when the peak of the light curve is reached, and $t_E$ is the characteristic {\it event timescale}, which is the time it takes for the source to move with respect to the lens by one Einstein ring~radius. }

The impact parameter may be determined at any time from the magnification by rewriting Equation (\ref{eqn:magn}) as
\begin{equation}
u = \sqrt{\frac{2A}{\sqrt{A^2-1}}-2}.
\end{equation}

Modeling single-lens event light curves usually involves three parameters that describe the shape of the curve, such as $u_0, t_0$ and $t_E$. Often, the baseline magnitude of the source is also included, which~is obtained when the source is far enough apart from the lens that lensing effects are not relevant.

\subsubsection{Microlensing Timescales}
Since $u_0$ and $t_0$ only depend on the trajectory of the source relative to the lens, the only parameter holding any physical information about the lensing system is $t_E$, because it depends on the distances between observer, lens and source, the mass of the lens, and the relative proper motion between source and lens, $\mu_{LS}$. We can express the event timescale as
\begin{equation}
t_E = \frac{\theta_E}{\mu_{LS}}. %\approx 0.214\mathrm{years} \left(\frac{M_L}{M_{\odot}}\right)^{1/2} \left(\frac{D_L}{10\mathrm{kpc}}\right)^{1/2}  \left(\frac{D_{LS}}{D_S}\right)^{1/2}  \left(\frac{200\mathrm{km/s}}{v_{\perp}}\right),
\end{equation}
%where $v_{\perp}$ is the relative lens-source transverse velocity on the lens plane.

If we use Equation (\ref{eqn:theta}) and assume a typical value for the proper motion  $\mu_{LS}\sim 15 \mu{\mathrm{as/d}}$, then for lenses of about one Earth-mass $t_E$ is of the order of hours, whereas in the case of Solar-mass objects the lensing effect can last for a few months \citep{2005MNRAS.362..945W,2015ApJS..216...12W,2016MNRAS.457.1320T}.

\subsubsection{The Optical Depth}
\label{sec:opdepth}
\textls[-15]{The optical depth to gravitational microlensing is the probability that a given star at a specific instant in time has a magnification caused by gravitational lensing that exceeds 1.34 \citep{1986ApJ...301..503P}, meaning that the source lies inside the Einstein ring of a lens. To evaluate it for a given source at distance $D_S$, we~need to consider all potential lenses lying along the line of sight to that source. The optical depth is then the integral over the number density of lenses multiplied by the area enclosed by their Einstein rings:}
\begin{equation}
\tau = \int_0^{D_S} \frac{4 \pi G \rho(D_L)}{c^2} \frac{D_L D_{LS}}{D_S} dD_L,
\end{equation}
where $\rho(D_L)$ is the average mass density of lenses at $D_L$.

The mass distribution depends on the line of sight considered so an accurate estimate of the optical depth requires the use of a Galactic model and the evaluation of the integral over multiple directions. Note however that the optical depth does not depend on the mass spectrum of lenses, but~only on the mass density along the line of sight.

To better understand the concept of optical depth, one can imagine an ensemble of point lenses randomly distributed in the sky, each lens surrounded by a small circle corresponding to its angular Einstein radius. Then, adding all those areas up gives $\tau$, the fraction of the sky they cover up.

How is that related with the distribution of event durations and the event rate? Continuing with the example above, each of these lenses is moving with some three-dimensional velocity in space. Even~though some fraction of them moves relatively fast, they will not all generate microlensing events with short timescales when they encounter a source star. This is because it is the velocity vector projected on the plane of the sky that affects the measured event duration, and if the magnitude of the projected vector is small, then the event duration will be long. On the other hand, slow-moving lenses will always generate long-timescale events. This means that the event duration distribution will be broad and have tails of relatively long and relatively short events. 

The event rate is the expected number of microlensing events $N$ if $n$ sources are monitored over a time interval $\Delta t$. A simple expression may be obtained under the assumption that all events have the same timescale $t_E$. Then $N=(2n\tau\Delta t)/(\pi t_E)$ \citep{1996ARA&A..34..419P}. A detailed analysis and the derivation of the full expression that takes into account the distribution of event timescales may be found in \citet{1996ApJ...473...57M}.

The first observational evaluation of the optical depth toward the Galactic bulge came from the OGLE survey in 1994 \citep{1994AcA.44.165U}, where it was estimated to be in excess of $3.3 \pm 1.2 \times 10^{-6}$. By 2005, the MACHO survey \citep{2005ApJ...631..879P,2000ApJ...541..734A} had refined this value to $\tau \sim 2.17^{+0.47}_{-0.38} \times 10^{-6}$. More recent estimates by \citet{2016ApJ...827..139S,2013ApJ...778..150S,2010GReGr..42.2047M} and \citet{2008ApJ...689.1078R}, find similar but slightly lower values. This means that, when looking from the Earth towards the Galactic bulge, only~about one or two stars in every million will be microlensed at any given time.

\subsection{Higher-Order Effects}
\label{sec:second_order1}
Most microlensing cases are adequately described by the standard point source-point lens model we outlined in the previous sections, but that is not always the case. Other, higher-order, effects may reveal themselves by perturbing the light curve of a microlensing event. Some of these increase the uncertainties in the measured parameters, whereas others may be used to extract extra information about the physical properties of the lensing system. 

\subsubsection{Blending}
\label{sec:blend}
Since microlensing events are such rare occurrences, surveys direct their observing efforts towards those regions of the sky with the highest number of stars. In practice, this means looking towards the very crowded stellar fields in the direction of the Galactic bulge, where stars appear so numerous that it often becomes difficult to disentangle them. However, it must be noted that the detection efficiency drops rapidly when crowding becomes so extreme that individual stars are no longer resolvable, such~as when observing microlensing events in other galaxies \citep{2007A&A...462..895I,2010MNRAS.404..604T,2010GReGr..42.2101C}.

Regular {\it blending} refers to the case when two or more stars overlap on the same image to such an extent that the software that performs the photometry cannot identify them individually, but rather as a single blurred object. Blending is a problem because it dilutes the true microlensing signal, which~is now `buried' in the light of all other stars contributing to the blend. As a result, if one forgets to account for blending, the event appears less magnified than it actually is, and the measured event duration is shorter since blended events spend less time above a given threshold. 

Taking blending into account involves adding a constant background to the time-variable flux received from the microlensing event, so that the total observed flux at any time is
\begin{equation}
F(t) = F_S A(t) + F_B,
\end{equation}
where $F_S$ is the source flux in the absence of lensing, $A(t)$ the magnification at time $t$, and $F_B$ the total blending flux. For detailed studies of how blending affects the determination of microlensing parameters, see \citet{2010ApJ...720..409H,2007MNRAS.380..805S,2006ApJ...640..299T} and \citet{1999MNRAS.309..373H}.

It is sometimes possible to constrain the contribution of the blend by obtaining high-resolution images, either with adaptive optic systems on the ground or from space with a telescope like the Hubble Space Telescope (HST). Given that the typical lens-source proper motion is of the order of \mbox{$\sim$5 mas/year}, this can only be done about a decade after the event was originally observed \citep{2016ApJ...824...83B,2017AJ....154....3K,2018AJ....155...78B}.

\subsubsection{Parallax}
\label{sec:parallax}
So far we have assumed that the relative proper motion between a lens and a source star can be approximated by a straight line, but the trajectory may well be more complicated. The parallax effect can be measured from the Earth or from space. It was first predicted for quasar microlensing in 1986~\citep{1986Natur.324..126G}, and then evaluated for stellar microlensing in 1992 \citep{1992ApJ...392..442G}. The first observational confirmation came in 1995 \citep{1995ApJ...454L.125A} by the MACHO survey.

\textls[-5]{In ground-based microlensing observations, the parallax effect consists of a subtle long-term distortion to the standard microlens light curve which arises from the orbital motion of the Earth around the Sun. Parallax is a vector with two components in the East and North directions, $\pi_{E,\mathrm{East}}$ and $\pi_{E,\mathrm{North}}$, following the Geocentric formalism of \citet{2004ApJ...606..319G}. The effect becomes particularly noticeable when the time scale of an event is some non-negligible fraction of the Earth's orbital period, i.e., when the event lasts for a few months. It is sometimes referred to as `orbital parallax', and may be expressed as }
\begin{equation}
\pi_E = \frac{\pi_{L}-\pi_{S}}{\theta_E} =
\frac{\theta_E}{\kappa M_L}; \quad
%t_E = \frac{\theta_E}{\mu_{\rm geo}}; \quad
%\theta_E^2 = \kappa M\pi_{\rm rel}; \quad
\kappa \equiv \frac{4 G}{c^2 {\rm AU}} \simeq 8.1 \frac{\rm mas}{M_{\odot}},
\end{equation}
where $\pi_{L}$ and $\pi_{S}$ are respectively the lens and source parallaxes, $M_L$ is the mass of the lens, $\theta_E$ the angular Einstein radius and $\kappa$ a constant \citep{2013ApJ...779L..28G}. 

Parallax is significant because for those microlensing events where both $\pi_E$ and $\theta_E$ can be measured, it is possible to determine the mass of the lens and the its distance from the Earth with great accuracy. The distance to the lens is given by
\begin{equation}
D_L = \frac{{\rm AU}}{\pi_E\theta_E + \pi_S},
\end{equation}
where $\pi_S= {\rm AU}/D_S$ is the parallax of the source star, which is usually known.

The problem is that most microlensing events last only for a few weeks, so the effect of `orbital parallax' is in most cases negligible, unless the magnification gradient is large with the source passing near the caustics (see Section \ref{sec:caust}). It is also possible that simultaneous observations from different observatories on the Earth may record small displacements to the event light curve due to the difference in perspectives. This is called `terrestrial parallax'.

`Satellite parallax', on the other hand, relies on having two observers, one on the Earth and a satellite in Solar orbit, observe the same microlensing event simultaneously. Since their points of view are slightly shifted with one another, the time when the event reaches its peak brightness and the perceived magnification will be different for each observer. In microlensing parlance, their $t_0$ and $A_0$ values will not be the same. \citet{2015ApJ...804...20C} and \citet{2015ApJ...802...76Y} first demonstrated this using simultaneous ground and space-based observations with the Spitzer satellite.

A source with a binary companion will also experience accelerated motion which can affect a light curve in a way analogous to ground-based parallax. This effect is called `xallarap' (parallax spelled backwards) and if the orbit of the source matches the reflex motion of the Earth around the Sun, then~the two effects manifest identically \citep{1998PhR...307...97B}.

Finally, `diurnal parallax' is a one-day modulation resulting from the rotation of the Earth around its axis \cite{2002ApJ...572..521A}.

\subsubsection{Finite Source}
\label{sec:finite}
Stars have a finite size but up to now we treated the source as point-like in the calculations. For~the most part this approximation is sufficient for single-lens events, but it breaks down when the impact parameter, i.e., the minimum distance between the lens and the source star, is comparable to or smaller than the size of the source (as a rule of thumb when $u_0 \lesssim 3 \rho_S$, where $\rho_S$ is the angular size of the source in units of $\theta_E$). To obtain the magnification for a source of finite size, one needs to integrate over the area of the source \citep{2010MNRAS.407.1597L,2009ApJ...695..200L,2007MNRAS.377.1679D}, or integrate over the source boundary using Green's theorem \citep{1998A&A...333L..79D}. It is worth mentioning that the source can have a brightness profile and might not be circular in projection.

When finite source effects are detected in the light curve on an event, they can be used to calculate the lens-source relative proper motion and evaluate the Einstein radius of the lens, since $\theta_E=\theta_S/\rho_S$. $\theta_S$ and $\rho_S$ are the angular radius of the source star and normalized source radius respectively. The former can be obtained from determining the spectral type of the source star, while the latter is evaluated during modeling of the event light curve. This is what is required for an accurate determination of the mass of the lens, as previously described in Section \ref{sec:parallax}.

\subsubsection{Binary Source}
The case of a binary source is simple, as it generates a linear sum of two single-lens light curves. Naturally, since the two stellar components may have different luminosities and colors, the resulting light curve may well be chromatic. If the distance between the two source stars is large and the lens trajectory is along the line joining the two, we may have the perception of two independent microlensing events separated by a few months or years, which are both produced by the same lens. Of course, if the binary source is a gravitationally bound system, it is possible that orbital motion effects may also be detectable.

The effect was first mentioned in a paper by \citet{1992ApJ...397..362G} in 1992. Some years later \mbox{\citet{1998ApJ...506..533G}} pointed out that there exist a subset of binary source events than can mimic the main features of planetary microlensing and thus masquerade as planetary events, contaminating the statistical sample. Fortunately, this degeneracy can often be broken by dense observations during the time of the perturbation and by using color information. \citet{2002ApJ...564.1015H} pointed out that if a lensing event is observed astrometrically, it is possible to unambiguously break the degeneracy.

Binary source events are not as commonly detected as originally predicted (\citet{1992ApJ...397..362G} estimated that $\sim$10\% of all events should show binary source features). This apparent paucity was explained by \citet{1998A&A...333..893D} and \citet{1998MNRAS.301..231H}; they are often mistaken for single source events.

All aforementioned higher-order effects have been detected and characterized in hundreds of microlensing light curves. The ways in which they affect or dilute the microlensing signal are generally well understood. Sometimes their presence can even be used to break degeneracies between the estimated parameters of the system, and especially in the cases of finite-source and parallax, to~determine the mass and distance to the lens.

\subsection{Binary Lens}
We now turn to the case when the lens is not a single body, but two: a binary star or a star with a planet. This is called `binary lensing', and the light curves it produces can show extraordinarily diverse morphologies \citep{2015MNRAS.450.1565L}.

\begin{figure}[H]
\centering
\includegraphics[width=14cm]{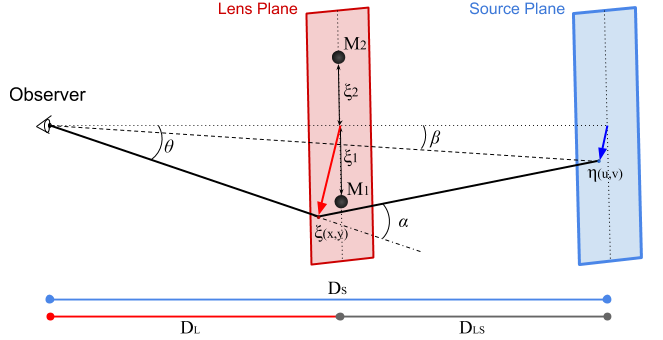}
\caption{The geometry of gravitational microlensing by a binary lens. The two components of the lens are at positions $\xi_1$ and $\xi_2$ on the lens plane, and the origin is chosen to be in the middle of the line joining the two masses. The distances from the observer to the lens and the source are $D_L$ and $D_S$, respectively. The distance between lens and source is $D_{LS}$. A ray of light originating from the source plane at point $\eta(u,v)$ at an angular position $\beta$, is gravitationally deflected by an angle $\alpha$ on its way to the observer. The position it hits the lens plane is denoted by $\xi(x,y)$, at an angular position $\theta$.}
\label{fig:binary_lens_side}
\end{figure}

Let us define our lens as two point masses $M_1$ and $M_2$ at a distance $D_L$ from the observer (see~Figure \ref{fig:binary_lens_side}). On the lens plane, we choose a coordinate system ($x,y$), where the $x$-axis passes through both masses and the origin is defined as the midpoint of the line joining them. $D_S$ and $D_{LS}$ are the observer-source and lens-source distances respectively. We define another coordinate system ($u,v$) on the source plane, which is parallel to ($x,y$), and whose origin lies on the point of intersection with the optical axis.

For any light ray, the deflection angle due to a single lensing mass is given by Equation (\ref{eqn:def_ang}). Assuming geometrically-thin lenses, when more than one lensing masses are involved, the total deflection angle is the vector sum of all individual deflections \citep{1992grle.book.....S,1973ApJ...185..747B}. When the contributions of each deflecting mass are added up, we can write the deflection angle for such a composite lens as
\begin{equation}
\label{eqn:nmass_lens}
\vec{\hat{\alpha}}(\vec\xi) = \frac{4G}{c^2} \sum_i M_i \frac{\vec\xi-\vec\xi_i}{\left|\vec\xi-\vec\xi_i\right|^2},
\end{equation}
where $\vec\xi$ is the position of the light ray on the lens plane, and $\vec\xi_i$ that of the mass $M_i$.

For the binary-lens case, this reduces to
\begin{equation}
\label{eqn:binary_lens}
\vec{\hat{\alpha}}(\vec\xi) = \frac{4G}{c^2} \left[M_1\frac{\vec\xi-\vec\xi_1}{\left|\vec\xi-\vec\xi_1\right|^2} + M_2\frac{\vec\xi-\vec\xi_2}{\left|\vec\xi-\vec\xi_2\right|^2}\right],
\end{equation}
where $\vec\xi_1,\vec\xi_2$ are lens-plane vectors pointing to $M_1,M_2$ respectively.  

The {\it binary-lens equation} can be written as
\begin{equation}
\label{eqn:bin_len}
\vec\eta = \vec\xi \frac{D_S}{D_L} - D_{LS}\vec{\hat{\alpha}}(\vec\xi),
\end{equation}
where the light ray originating from a source at $\eta(u,v)$ hits the lens plane at image position $\xi(x,y)$. As~before, $D_S, D_L$ and $D_{LS}$ denote the observer-source, observer-lens and lens-source distances.

Following \citet{1986A&A...164..237S}, we can use the definitions $r_E = \theta_E D_L$, $\vec{z}=\vec{\xi}/r_E$, \mbox{$\vec{w}=(D_L/D_S) \cdot \vec{\eta}/r_E$}, and express the masses as fractions of the total lens mass, $\epsilon_1 = M_1/M_L$, \mbox{$\epsilon_2 = M_2/M_L$}, so that Equation (\ref{eqn:bin_len}) can be written as
\begin{equation}
\label{eqn:bin_len3}
\vec{w} = \vec{z} - \epsilon_1 \frac{\vec{z}-\vec{z_1}}{|\vec{z}-\vec{z_1}|^2} - \epsilon_2 \frac{\vec{z}-\vec{z_2}}{|\vec{z}-\vec{z_2}|^2},
\end{equation}
with $\vec{w}(u,v)$ on the source plane mapping to $\vec{z}(x,y)$ on the lens plane. In other words, through Equation (\ref{eqn:bin_len3}), we can find the source point $\vec{w}(u,v)$ at which a light ray which hits the lens plane at $\vec{z}(x,y)$ originates. 

\citet{1990A&A...236..311W} rewrote Equation (\ref{eqn:bin_len3}) using complex notation, which makes it more convenient to use. If $w=u+iv$ and $z=x+iy$ are now complex numbers and $\bar{w},\bar{z}$ their complex conjugates, we can obtain $w$ as a function of $z$ and $\bar{z}$:
\begin{equation}
\label{eqn:bin_len_complex}
w = z - \frac{\epsilon_1}{\bar{z}-\bar{z_1}} - \frac{\epsilon_2}{\bar{z}-\bar{z_2}}.
\end{equation}

Here, we have adopted the notation of \citet{1997ApJ...484...63R}, where the general case for $N$ lensing bodies is~described.

We are usually more interested in knowing where a source, located at $w(u,v)$, can be seen on the lens plane; thus we need to invert Equation (\ref{eqn:bin_len_complex}). This is non-trivial as it is of fifth degree in $z$ and the inversion cannot be done analytically for arbitrary $w$. However, the roots may be found using standard numerical recipes, such as the ZROOTS routine from \citet{2002nrca.book.....P}, or the optimized algorithm of \citet{2012arXiv1203.1034S}.

\subsection{The Magnification and Image Positions for the Binary Lens}
As in the single lens case, the magnification of an image $i$ is obtained by taking the ratio of the flux density of the lensed image and the flux density of the unlensed source. This is determined by evaluating the Jacobian determinant, $J$, of the mapping by the binary-lens equation:
\begin{equation}
\label{eqn:jacob}
A_i = |J|^{-1}; \quad
J = 1 - \left|\frac{\partial w}{\partial \bar{z}}\right|^2; \quad 
\frac{\partial w}{\partial \bar{z}} = \frac{\epsilon_1}{(\bar{z}-\bar{z_1})^2} + \frac{\epsilon_2}{(\bar{z}-\bar{z_2})^2}.
\end{equation}

The need to evaluate the Jacobian arises because we want to know how an infinitesimal area element in the lens plane is distorted during mapping through the lens equation.

The total magnification is simply given by the sum of the magnifications of all images:
\begin{equation}
A = \sum_{i=1}^n A_i = \sum_{i=1}^n |J|^{-1}.
\end{equation}

\subsection{Critical Curves and Caustics}
\label{sec:caust}
\begin{figure}[H]
\centering
\includegraphics[width=14cm]{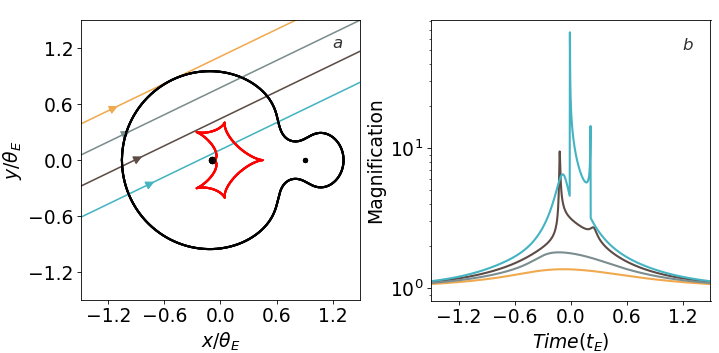}
\caption{{(\textbf{a})} Critical curves (black) and caustics (red) of a binary lens with $s$ = 1, $q$ = 0.1, assuming a point source. The origin is the center of mass of the system. When the lens is composed of two massive objects, the images of the source generated by the lensing effect are no longer two, but either three of five, depending on the location of the source relative to the caustics. For each of the four source trajectories considered, two new images get created or destroyed whenever the source crosses the caustic, producing the sudden `jumps' in magnification seen in panel {\bf b}. The positions of the two lensing masses are marked by the black dots and the $x$ and $y$-axes mark distances on the lens plane in units of $\theta_E$; {(\textbf{b})} Corresponding light curves for each of the trajectories shown in panel {\bf a}. Note the absence of sudden `jumps' in magnification for the trajectories shown in orange and gray, where the source remains far from the caustics at all times.}
\label{fig:binary_lens_caustics}
\end{figure}

Certain values of $\bar{z}$ in Equation (\ref{eqn:jacob}) will make the Jacobian determinant vanish. These positions trace out closed curves on the lens plane called {\it critical curves}. Mapping those through the lens equation onto the source plane, one obtains another set of closed curves called {\it caustics}. Critical curves and caustics are of fundamental importance in understanding microlensing by two or more lensing bodies. That is, at these positions the magnification diverges and reaches infinity for a point source (\mbox{see Figure~\ref{fig:binary_lens_caustics}}). In reality however, because real sources are extended and the magnification is a weighted mean over the source, the magnification is always finite.

The significance of the caustics becomes easier to appreciate when one considers that whenever the source crosses a caustic, the number of images changes by two, leading to abrupt `jumps' in magnification. When the source is outside the area enclosed by the caustic curve, the number of images is always three. As the source crosses the caustic, two more images are produced at diametrically opposite locations \citep{1996ARA&A..34..419P}, and for as long as the source remains enclosed within the caustic structure, the total number of images is always five. Generalizing, the total number of images generated by a lens consisting of $N$ lensing masses correspond to the solutions of the complex polynomial obtained from inverting the lens equation and therefore cannot exceed $N^2 + 1$ \citep{1997ApJ...484...63R}. In fact, not all solutions of the complex polynomial correspond to real images; some are spurious. \citet{2003astro.ph..5166R} demonstrated that at most $5(N-1)$ real images will be generated. As in the case of a single lensing mass, the image separations are too small to resolve them individually, but binary lens light curves exhibit much more complicated structures.

\begin{figure}[H]
\centering
\includegraphics[width=12cm]{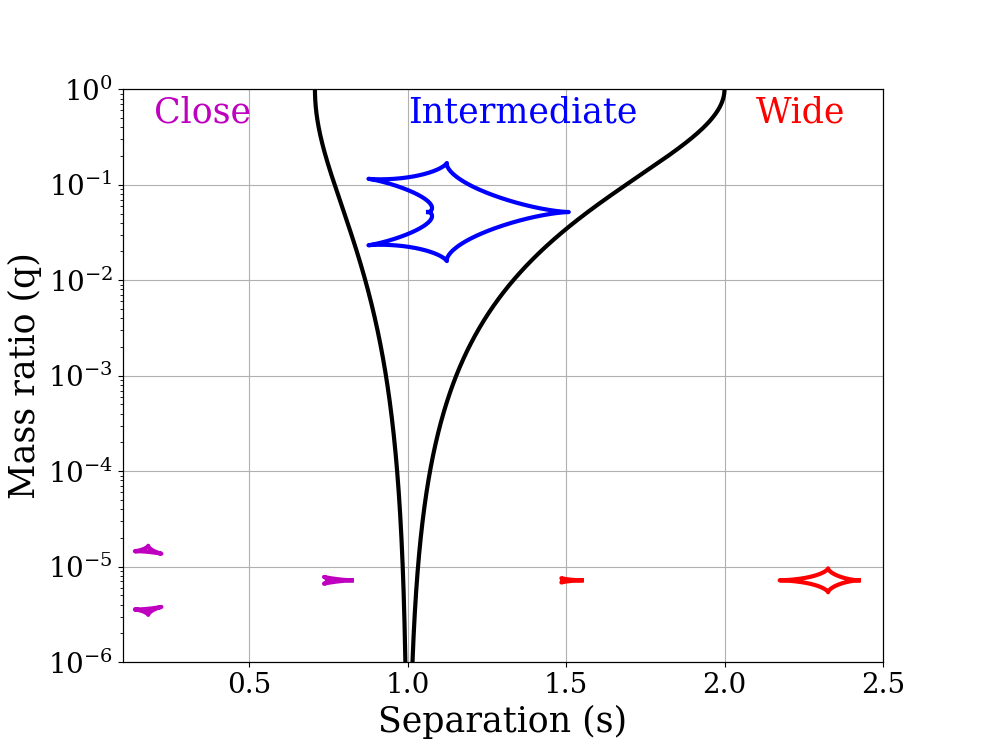}
\caption{Three possible caustic topologies: close, intermediate/resonant and wide, with black lines separating the three regimes. A close binary (in the example plotted here $s=0.75$, $q=10^{-2}$) will produce a central four-cusped caustic and two smaller three-cusped symmetrical ones. An intermediate resonant binary ($s=1$, $q=10^{-2}$) has a single large six-cusped central caustic, while a wide binary ($s=1.5$, $q=10^{-2}$) has a central and an isolated secondary caustic, both featuring four cusps.}
\label{fig:caustic_topologies}
\end{figure}

The shape of the caustics is very sensitive to the mass ratio between the two components of the lens, $q$, and different mass ratios generate different caustic structures. The shape of the caustics is also sensitive to the angular separation, $s$ (in units of $\theta_E$), between the two components of the lens, with larger separations `stretching' the caustics along the binary axis, while beyond a certain point the caustics split up. Different values of $q$ and $s$ will produce caustics that belong to one of three possible topologies: close, intermediate/resonant and wide \citep{1986A&A...164..237S,1993A&A...268..453E,2006ApJ...638.1080H,2008A&A...491..587C,2010exop.book...79G,2016ASSL..428.....B}. These are illustrated in Figure~\ref{fig:caustic_topologies}. Close topologies are characterized by three separate caustic curves: a central caustic and two smaller symmetrical ones on either side of the binary axis. These are usually called planetary caustics since they are associated with the smaller of the two masses. In wide topologies, there is a central caustic and an isolated secondary caustic that appears along the binary axis. Finally, an intermediate/resonant topology produces a single large central caustic. The shape of the observed light curve will then depend on exactly how close the trajectory of the source brings it to these curves.

As can be seen in Figure \ref{fig:binary_lens_caustics}, when the source trajectory is far from the caustics, the resulting light curve is similar to the single lens case. Only those trajectories that pass close to or cross the caustics will produce significant deviations. Observations of microlensing events only provide a light curve, from which the size and structure of the caustics must be inferred. Provided the photometry is good, the exact morphology of the light curve deviations can be used to tell whether a microlensing event is caused by a star-star binary lens or by a star and its planet. The duration of a deviation scales roughly as $\Delta t_p \sim \sqrt{q}t_E$, so for a typical event timescale $t_E = 30$ days, a Jupiter-Sun equivalent (mass ratio $q=10^{-3}$) would produce a deviation lasting for about a day, whereas the deviation caused by an Earth-Sun equivalent ($q=3 \times 10^{-6}$) would only last for about two hours.

An alternative way to think about light curve deviations is to consider what happens to the images while the lensing event is ongoing. Let us for example consider a binary lens composed of a star hosting a planet. The planet will only be revealed when one of the images generated by the lensing event happens to sweep past its location on the lens plane. The gravity of the planet will then act as a mini-lens, further perturbing the image and producing a `planetary anomaly' in the light curve.

\subsection{Binary Light Curve Degeneracies}
As we saw in Section \ref{sec:pspl_mag}, three parameters are enough to describe the shape of a point-source, point-lens light curve: the minimum impact parameter, $u_0$, the time of maximum magnification, $t_0$, and the event timescale, $t_E$. Binary lenses require three additional parameters: the angle $\alpha$ at which the source trajectory crosses the binary axis, the mass ratio between the two components, $q$, and the separation between them, $s$, projected on the lens plane. If the source at any point approaches the caustics, its angular size, $\rho_S=\theta_S/\theta_E$, must also be included as an extra parameter, bringing the total number of parameters required to describe the shape of the binary light curve to seven.

Determining these parameters through light curve modeling is sometimes complicated due to a number of well-studied degeneracies \citep{1997ApJ...486...85G,1999A&A...341..943D,2009MNRAS.393..816D}. A binary lens where one of the components is a planet, is subject to two discrete degeneracies. The first relates to which image the planet is perturbing; Is it the major image outside the Einstein ring or the minor image inside? Since these result in different light curve perturbations, this degeneracy can easily be broken for the majority of cases. The second degeneracy has to do with whether the planet lies closer to or further from the star than the position of the image it is perturbing. This is harder to break but does not affect the determination of the mass ratio. In addition to these discrete degeneracies, there is a continuous degeneracy arising from finite-source effects being misinterpreted as a larger mass ratio, because both of these are related to the duration of the planetary perturbation. If, for example, the size of the source is larger than the Einstein ring of the planet, then the duration of the perturbation will be determined by the crossing time of the source, not the Einstein ring of the planet. Another type of degeneracy, whose origin lies in the lens equation itself, exists between close and wide binary lenses \citep{1999A&A...349..108D}. Gaps in the observations, especially when the source is traversing the caustics, may also permit multiple degenerate model solutions since the shape of the light curve during the anomaly is only loosely constrained \citep{2018AcA....68...43S}.

In general, dense and precise sampling of the light curve with high-cadence observations will place tight constraints on possible degenerate solutions. As we shall see next, when higher-order effects are detected in the light curve, it is often possible to break degeneracies and arrive at unique solutions for the physical parameters of the lensing system.

\subsection{Higher-Order Effects for Binary Lenses: Orbital Motion}
\label{sec:second_order2}
All higher-order effects that apply to single lenses (see Section \ref{sec:second_order1}), can also affect binary lenses. In fact, the finite size of the source $\rho_S$ is even more important in binary lenses since its effects need to be taken into account whenever the source approaches a caustic, whereas for single lens events it is rarely required. Similarly, parallax effects are much more frequently detected in binary than in single lens events. In addition to these, it is sometimes possible in binary lens events to detect evidence of orbital motion between the lens components.

Typically, microlensing events last only for a few weeks, so the positions of the two masses can be considered fixed on the lens plane. This means that our observations will only get a `snapshot' of the system. However, when the duration of the microlensing event is some non-negligible fraction of the orbital period of the binary and the source trajectory passes close to the caustics, the orbital motion of the two bodies comprising the lens will leave its traces in the event light curve \citep{1998A&A...329..361D,2000ApJ...534..894A}.

Orbital motion manifests itself by changing the shape of the caustics with time and rotating them, but since only a small part of the orbit is covered during the microlensing event, it is usually only a small effect. However, the effects can be particularly dramatic on the planetary caustics for close binaries. To first order approximation, orbital motion can be accounted for by introducing two more parameters to the binary lens model: the rate of change of separation between the lens components $ds/dt$ (projected on the lens plane), and the angular rotation rate $d\alpha/dt$ (relative to the source trajectory). It is worth noting that parallax and orbital motion effects are partly degenerate but can often be disentangled with careful observations \citep{2012ApJ...754...73B,2013ApJ...778..134P}. Remarkably, for a few systems it has been possible to derive complete Keplerian solutions for the orbits \citep{2011ApJ...735...85S,2012ApJ...755...91S}. 

\subsection{Finding Planets}
Planetary microlensing events are just binary lenses where the mass ratio between the two components of the lens is very small. Strictly speaking, they can be (and indeed have been) found in more complex systems involving three or more lensing objects, but these are much rarer occurrences. Because the mass of the planet is so small compared to the host star, their light curves for the most part resemble those of single lenses. The planet will reveal itself when one of the images of the source generated by the lensing effect happens to `sweep' past its location. This will produce brief binary-lens type perturbations on the light curve, usually referred to as `planetary anomalies', which after careful observations and analysis can be used to derive the characteristics of the system. Since the images appear close to the Einstein ring of the lens, that is where the probability of detecting a planet, if it is there, is highest. For typical lens and source distances, this corresponds to a physical distance from the host star of about 1--10 AU \citep{2002ApJ...566..463G,2003MNRAS.343.1131T,2010ApJ...720.1073G}. Although the amplitude of the planetary signal can be large irrespective of the mass of the planet, the detection probability scales approximately as $q^{1/2}$ in the vicinity of the planetary caustics or as $q$ close to the central caustic \citep{1998ApJ...500...37G,2001ApJ...552..889P}. 

The analysis of microlensing event light curves featuring perturbations due to the presence of multiple planets can be a daunting task requiring optimization over a large parameter space. Multi-planet systems form several disconnected sets of caustics; there is a 'central caustic' located very close to the primary lens and multiple 'planetary caustics' further away. Perturbations related to the source interacting with the central caustic appear near the peak of the light curve, whereas those related to interactions with the planetary caustic manifest at the wings on either side. Provided the deviations induced by the individual planets do not occur at similar places, most cases of planetary caustic perturbations can be adequately addressed using the binary-superposition approximation, whereby the perturbations are treated as the sum of the individual perturbations caused by each planet. These~can be investigated separately using standard binary lens analysis methods and repeated for each planet \citep{2001MNRAS.328..986H,2005ApJ...629.1102H}. Although the approximation still holds for many cases of central caustic perturbations, the situation becomes more complex since these always occur in the same region regardless of the number of planets \citep{2002MNRAS.335..159R}. This means that it can be difficult to isolate the individual contribution of each planet in the pattern of anomalies observed in the light curve, especially when the separation between the planets is small. In general, the binary-superposition approximation is valid for $q \ll 1$ and $|s-1| \gg q$ and will provide a rough first estimation of the parameters of the event which can then be refined by performing the full analysis in a greatly narrowed-down parameter space.

This concludes the theory section. What ultimately determines the degree of success in detecting and characterizing planetary signals is the timeliness, quantity and quality of observations, and it is to this we turn to next. 

\section{Microlensing Observations in Practice}
What we actually observe during a microlensing event is an increase in the brightness of the source star as the lens approaches it on the plane of the sky, followed by a gradual dimming back to its normal brightness as the lens appears to move away (see Figure \ref{fig:single_lens}). The relative proper motions between the stars in the Galaxy produce microlensing events that typically last a few weeks to a few months. If the lensing star happens to host a planet, there is a chance that the planet itself may also perturb the light coming from the source star, producing brief but intense variations, or `anomalies', in~the event light curve, i.e., in the measurements of how the source brightness varies with time. These anomalies typically last for a few days in the case of Jupiter-mass planets and only for a few hours for Earth-mass planets, but the amplitude of the anomaly can be substantial in both cases. Planetary deviations are detected in $<1\%$ of all microlensing events discovered.

As mentioned in Section \ref{sec:whatis}, microlensing is a rare phenomenon; only about one star in every million undergoes microlensing at any given time in the Galaxy. Therefore, in order to maximize their chances of finding these elusive events, microlensing surveys have been targeting those regions of the sky with the highest density of stars per square degree; and this means looking in the direction of the Galactic bulge.

\subsection{Surveys and Follow-Up}
\textls[-20]{The first microlensing surveys were actually set up to investigate whether a significant fraction of the dark matter in the Milky Way halo was made up of massive compact objects (MACHOs) that emit little or no radiation, like black holes, neutron stars and brown dwarfs \citep{1992AcA....42..253U,1995Msngr..80...31A, 1996ApJ...471..774A, 1997A&A...324L..69R}, but little evidence was found to support this hypothesis. It wasn't until \citet{1991ApJ...374L..37M} and \mbox{\citet{1992ApJ...396..104G}} pointed out that planets could be discovered in this manner that efforts started to move in that direction.}

In the early to mid '90s, microlensing survey telescopes only covered what by today's standards is but a small area of the sky, of the order of $\sim$1 degree, finding few microlensing events. For~example, the~OGLE-I survey that lasted from 1992 until 1995 discovered 19 microlensing events, which at the time was a resounding success. The pivotal point came with the introduction of the OGLE early-warning system in 1994, which allowed early detection of on-going microlensing events in their data stream~\citep{1994AcA....44..227U}. They decided to release this information publicly and promptly to the astronomical community, enabling follow-up observations from other telescopes, so that as much information about these one-off events could be obtained. This was necessary given that surveys at that time lacked the observing cadence required to constrain the parameters of the most interesting among the events they were discovering. 

This decision, soon emulated by other surveys, enabled small dedicated teams around the world to strongly contribute to the science \citep{1995astro.ph..8039P,1998ApJ...509..687A,2004ApJ...603..139Y,2007P&SS...55..582B,2009AN....330....4T,2010AN....331..671D}. Under this tacit agreement, surveys provided the alerts and daily monitoring necessary to discover the microlensing events, whereas follow-up teams concentrated their efforts on obtaining observations every few hours for a small subset of events that were either highly-magnified, which implied that the probability of discovering potential planetary companions was high, or that were already exhibiting anomalous features \citep{2007ApJ...661.1202H,2013MNRAS.431.2975A}. As pointed out by \citet{2002P&SS...50..299D}, the OGLE-III survey made the crucial difference for the efficiency of follow-up campaigns by providing a much larger number of useful events on brighter targets (a factor 5–8 as compared to OGLE-II). Some follow-up teams removed the human-decision element altogether and introduced automated processes that could detect and assess ongoing anomalous features in real time, relegating key observing decisions to software agents \citep{2002MNRAS.337...41T,2007MNRAS.380..792D,2008AN....329..248D}. 

This arrangement benefited all parties for a number of reasons: First, surveys possessed cameras with a wide field of view, allowing them to discover and issue alerts about ongoing microlensing events. This was something follow-up teams could not do due to limitations imposed by the designs of the telescopes, which generally had much smaller fields of view. Secondly, follow-up teams independently operated different telescopes around the world, allowing for near-continuous observations of a microlensing event, when surveys could only obtain observations when it was a clear night at the location of their telescope. This made it much more likely that all interesting features of an event could be observed. Thirdly, unlike surveys, follow-up teams had the option of tailoring the exposure times of their observations to the current brightness of the individual event they were monitoring, thereby maximizing the signal-to-noise ratio while avoiding saturation. 

One of the difficulties of this approach was combining disparate data sets, given that the telescopes and instrumentation used by each team had different characteristics. In order to achieve the required accuracy to detect planetary signals ($\le$1\% photometry), all data sets had to be carefully aligned and the noise properties of each evaluated \citep{2015ApJ...812..136B}. What's more, observations were reduced in real time and light curves made promptly available to the community so that putative signals could be independently assessed. Based on these `live' data, a quick decision could be made on whether individual observing schedules required adjustments. For many of the anomalous microlensing events observed during this period (roughly between 2005 to 2015), up to 20 telescopes around the world contributed observations, providing 24-h coverage of the light curve and independent confirmation of the anomalous features.

\subsection{Real-Time Modeling}
Precise modeling of individual microlensing events is a process that can last several months and is usually performed on computing clusters with hundreds of CPUs because a large parameter space needs to be thoroughly explored. The nature of anomalous features in the light curve of an event while the anomaly is still ongoing is very hard to interpret. Since the main feature containing information about the mass of the secondary component of the lens is the duration of the anomaly, it is often impossible to tell whether it is a planet, a brown dwarf, or a companion star causing it. As the event progresses and more observations are obtained, the likelihood of the different interpretations changes, but estimating exactly by how much requires the ability to model events in real-time as they are happening, and to constantly re-evaluate how well these models represent the data. This~was computationally prohibitively expensive until 2010 when, based on previous work in the field, \mbox{\citet{2010MNRAS.408.2188B}} came up with a conceptually simple method that could evaluate competing models in-real time and help follow-up teams decide whether to continue intensive observations or to reallocate their resources to a different target \citep{2010ApJ...723...81R,2012MNRAS.424..902B}.
Furthermore, the software product was made publicly available and could be run on an average laptop. This made a complicated process accessible to the entire community and this code is now at the core of most publicly available microlensing modeling software \citep{2017AJ....154..203B,2018arXiv180301003P,muLAn}.

\subsection{Second-Generation Surveys}
The OGLE and MOA microlensing surveys have gradually upgraded their instrumentation, installing new wide-field cameras on their telescopes, and can currently cover several square degrees on the sky with a single pointing \citep{2008ExA....22...51S,OGLE4cam,moa,ogle}. Hundreds of millions of stars are now monitored as often as every 10--30 min in the direction of the Galactic bulge, and this substantially improved survey sensitivity to planetary signals in the light curve. Follow-up observations from different longitudes are still useful for independently confirming the signals, as well as providing light curve coverage when the Galactic bulge is not visible from the survey sites, either due to inclement weather or during daytime \citep{2011ApJ...741...22M,2011ApJ...728..120M,2013A&A...552A..70K,2014ApJ...782...48T}. Survey efforts were greatly augmented in 2016, when the Korea Microlensing Telescope Network (KMTNet) commenced full operations with its three 1.6 m survey telescopes in Chile, Australia and South Africa, which were deployed and commissioned during 2015 \citep{2011SPIE.8151E..1BK,2014ApJ...794...52H,kmtnet}. The cameras on each of the KMTNet telescopes have a field of view of $\sim$4 square degrees and their $\sim$10 min cadence observing zone covers about 16 square degrees on the sky.

\subsection{From Digital Images to Light Curves}
Telescopes carrying out microlensing observations are equipped with CCD or EMCCD cameras that produce high-quality digital images. On a typical night during the microlensing season, lasting approximately from April to September each year, when the Galactic bulge is visible from the Southern hemisphere for six or more hours, dozens to hundreds of images will be obtained. The achievable photometric precision depends on the performance of the camera at the chosen optical wavelength, the observing conditions, the brightness of the target, the degree to which its light is blended with that of other nearby stars on the image, and the method used to extract the photometric measurements. 

All images undergo standard preliminary processing to remove noise due to the camera electronics and to correct for irregularities in the optical path, such as specks of dust on the optics \citep{1997pgca.book.....M,2006hca..book.....H}. Because telescope pointing is not perfect, all images are aligned to a common reference so that stars occupy the same pixel areas on every image. Photometry is performed by measuring the intensity of each star, i.e., its flux, at the given position. This is typically done in one of three ways: {\it aperture photometry}, {\it point-spread function (PSF) fitting}, or {\it difference imaging}. 

Aperture photometry is conceptually the simplest and involves summing up the flux contribution of each pixel within an aperture of fixed radius centered on the position of the star to be measured~\citep{1999ASPC..189...50M}. Comparison stars of constant brightness on the same image are also measured and used to remove artificial trends in the data. While this method works well when there are few isolated stars on the image, it is not designed to perform well in crowded field conditions, which are characteristic of microlensing observations. In crowded fields, such as those in the direction of the Galactic bulge, stellar profiles are too blended on the images and it is hard to disentangle the individual contributions to the total flux measured within the aperture.

PSF fitting photometry does better when there is a moderate degree of blending and is still used extensively for many astronomical projects. About 20--30 isolated stars are selected on an image and a scalable mathematical stellar profile is constructed by fitting a model to the image data. The model can be a sum of 2D Gaussian or Moffat functions. This PSF profile is then scaled to fit all stars identified on the image \citep{1999ASPC..189...56H}, and the volume under the profile corresponds to the total flux of the star. PSF-fitting was routinely used for the analysis of microlensing observations until \citet{1996AJ....112.2872T} and \citet{1998ApJ...503..325A} proposed the more efficient difference imaging method for measuring stellar flux variability in crowded fields.

\begin{figure}[H]
\centering
\includegraphics[width=14cm]{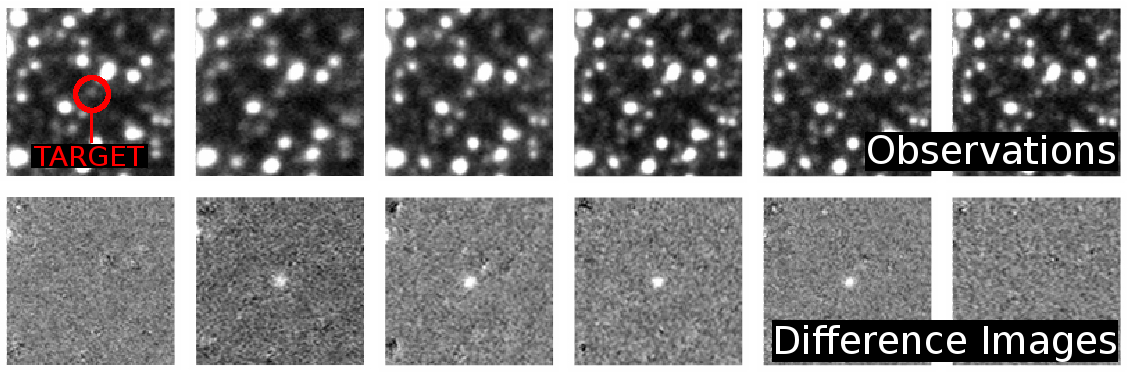}
\caption{[Top] 30$\times$30 pixel thumbnails of an image sequence with the microlensing target at the center. [Bottom] Residual difference images corresponding to each image above after subtraction of the reference image. The change in brightness of the microlensed star is hard to spot in the top row but is clearly visible in the bottom.}
\label{fig:diff_img}
\end{figure}

The technique of difference imaging involves the subtraction of all constant features from the images, using them to self-calibrate the photometry, and then using the difference images to measure the brightness fluctuations of a variable star, such as a microlensing target. Any number of observations taken under excellent observing conditions can be combined to produce a single reference image with sharp features. This reference is then adjusted to match the observing conditions of every other image of the target, taken at different times. The adjustment involves a convolution that shifts the coordinates of the stars to match small shifts in telescope pointing between the images and a `blurring' and scaling of stellar profiles in order to match different atmospheric seeing conditions and transparency. The~relevant transformations are applied to the reference, and it is then subtracted from each other image, removing all constant features and leaving behind only the signals of stars that have varied between different exposures. The brightness variations of microlensing events show up prominently on the residual images and can by measured by PSF fitting or aperture photometry (see Figure \ref{fig:diff_img}). Difference imaging is, on average, a factor of $\sim$2 more accurate than PSF fitting, and can even perform $\sim$7 times better for very faint targets \citep{1999A&A...343...10A}. The technique has been improved upon and extended in recent years \citep{2008MNRAS.386L..77B,2009MNRAS.397.2099A, 2013MNRAS.428.2275B} and is now the preferred choice for analyzing microlensing observations.

\section{Results from Microlensing}
Microlensing has detected more than 60 exoplanets to date (see Figure \ref{fig:sensitivities}), including two systems with multiple planets \citep{2008Sci...319..927G,2013ApJ...762L..28H}, a planet orbiting a double star system \citep{2016AJ....152..125B}, and the first possible detections of exo-moons \citep{2014ApJ...785..155B,2018AJ....155..259H}. The majority of these discoveries have been the result of close collaboration between survey and follow-up teams and, in most cases, the planetary signals have been confirmed in multiple data sets. The published parameters for all of these systems can be found in the \href{https://exoplanetarchive.ipac.caltech.edu/}{NASA exoplanet archive} \citep{exoarchive1} or the \href{http://exoplanet.eu/}{Extrasolar Planet Encyclopaedia} \citep{exoarchive2}, which also lists companions in the brown dwarf mass range. Both archives are regularly updated.

\begin{figure}[H]
\centering
\includegraphics[width=14cm]{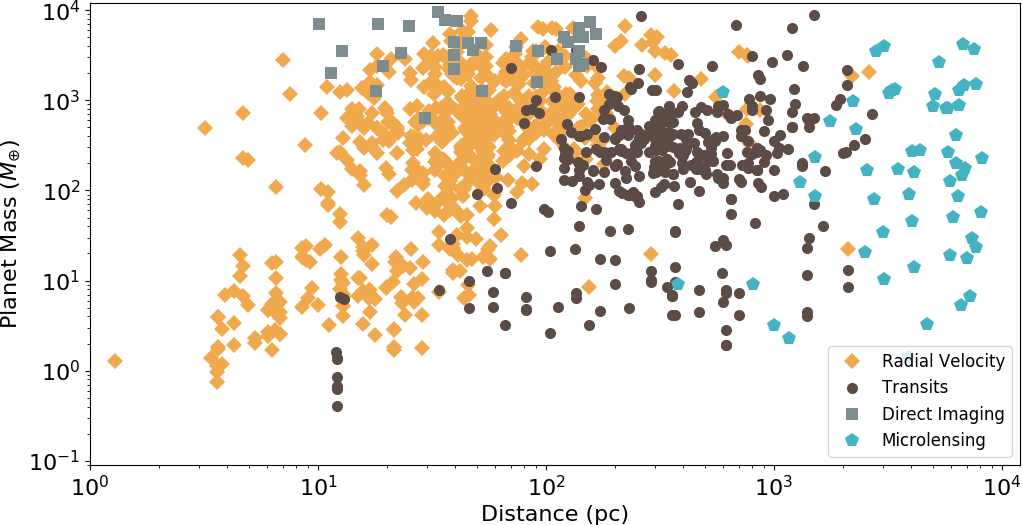}
\caption{Almost all known exoplanets lie within a few hundred parsecs from the Sun. The exception are planets found with microlensing, since most lens stars are at distances of several kilo-parsec. Microlensing is therefore the only technique capable of exploring the true Galactic population of planets. Stars closer to the Galactic center are generally more metal-rich than those in the halo, and this can have implications on the type of planetary systems they host.}
\label{fig:distances}
\end{figure}

These microlensing planets orbit their stars (predominantly K and M-dwarfs) at separations between 0.5 and 18 AU, a regime that remains largely unexplored by radial velocity and transit surveys, and their masses range from $\sim$1.4 $M_{\mathrm{Earth}}$ to $\sim$13 $M_{\mathrm{Jupiter}}$. Beyond $13 M_{\mathrm{Jupiter}}$, which is the approximate minimum mass required to start deuterium fusion, the distinction between planet and brown dwarf becomes difficult \citep{2011ApJ...727...57S}. Furthermore, it must be noted that all microlensing planets have been discovered at distances of several kilo-parsec away from the Solar system, along the line-of-sight to the Galactic center. They are much further away than the planets found by any other method (\mbox{see Figure \ref{fig:distances}}). This is particularly interesting because the Galaxy has a metallicity gradient \citep{2004ApJ...617.1115D}, and~a link has been established between stellar metallicity and the frequency of giant planets \citep{1997MNRAS.285..403G,2008PhST..130a4003V}. This~could mean that the statistical properties of the systems microlensing is discovering do not follow the same distributions as those found by other search methods, which probe only nearby stars. Thus, microlensing is currently the only way to explore and understand the true Galactic population of planets. At the current rate, only a few microlensing planets are discovered every year, so building a large enough statistical sample in order to address this question might take decades.

\subsection{Highlights}
It is instructive to consider two microlensing events of note and examine what has been learned from studying the morphology of their light curves.

\subsubsection{A Cold Super-Earth Orbiting an M-Dwarf Star}
The discovery of a planet in microlensing event OGLE-2005-BLG-390 in 2005 was noteworthy for three reasons \citep{2006Natur.439..437B}. First, at the time of publication, it was one of the lowest-mass planets known. Second, with a mass of $\sim$5.5 %^{+5.5}_{-2.7} 
$M_{\mathrm{Earth}}$ and a distance from its M-dwarf host of 
$\sim$2.6 %^{+1.5}_{-0.6}
  AU, which implied a surface temperature of  $\sim$50 K, it was the only planet known with a potentially solid surface composed of rock and ice. Third, the discovery gave credence to the idea that terrestrial-mass planets orbiting low-mass stars between 1 and 10 AU are far more common than Jupiter-mass planets, as predicted by the core-accretion model of planet formation \citep{2004ApJ...612L..73L}.

\begin{figure}[H]
\centering
\includegraphics[width=12cm]{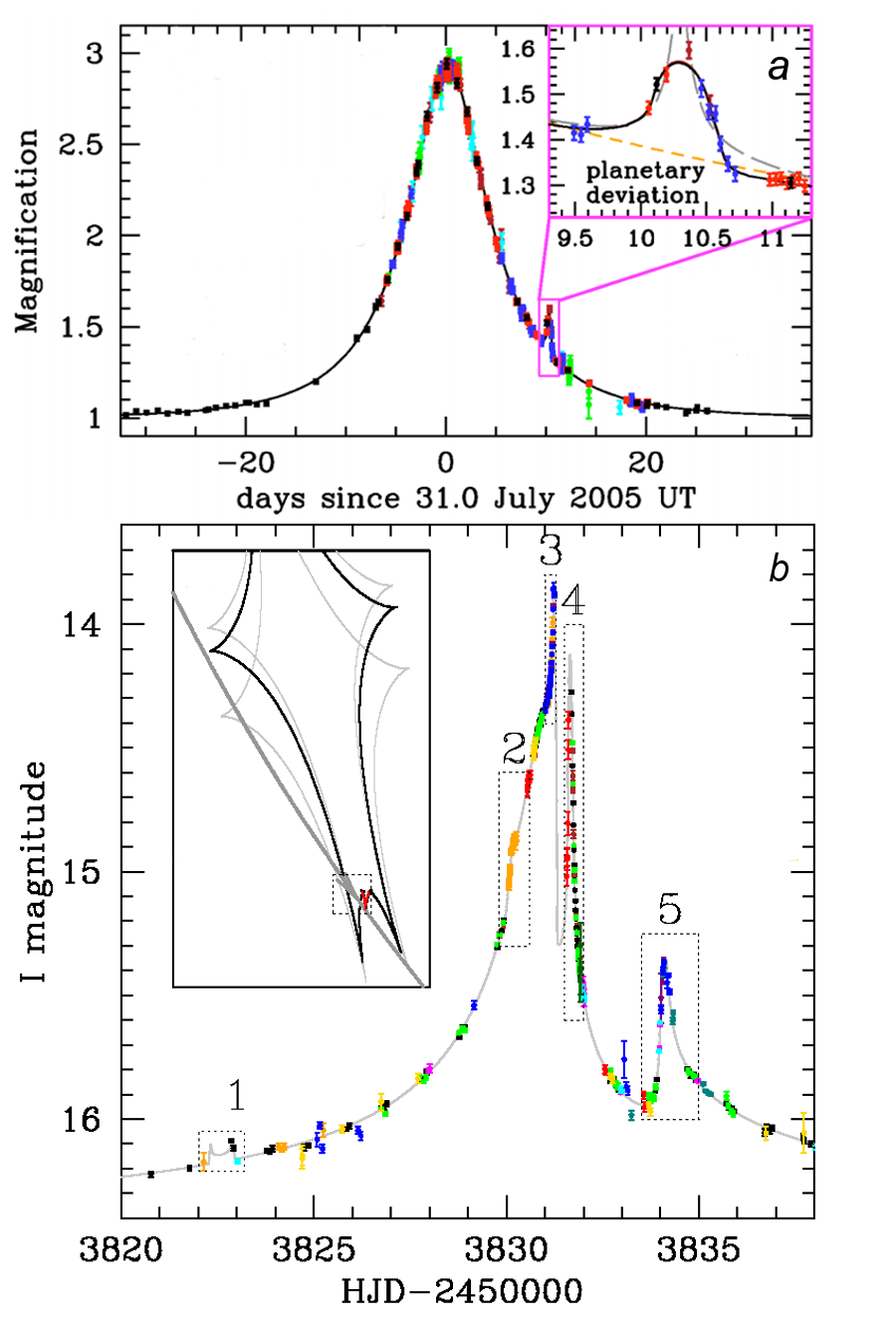}
\caption{{(\textbf{a})} The light curve of planetary microlensing event OGLE-2005-BLG-390. Six telescopes around the world participated in this discovery, with the individual contributions represented by the data points in different colors. The inset on the top right shows the planetary anomaly with the best-fit planetary model indicated by the solid black line. Single lens and binary source models, respectively represented by the orange and gray dashed lines, cannot account for the observed deviation. The~planet has a mass of $\sim$5.5 %^{+5.5}_{-2.7} 
$M_{\mathrm{Earth}}$ and lies at a distance of 
$\sim$2.6 %^{+1.5}_{-0.6}
 AU from its host star. [{\it Figure adapted with permission from \citet{2006Natur.439..437B}, Figure 1}]; {(\textbf{b})} The light curve of planetary microlensing event OGLE-2006-BLG-109. The event was observed from 11 different telescopes, represented by the data points in different colors. Five distinct anomalous features are apparent, produced by the source crossing the caustic of the two-planet system (shown in the inset on the left). The gray caustics show how the structure of the caustics appears at different times due to the orbital motion of the outer planet. The source trajectory is slightly curved due to the effect of annual parallax. [{\it Figure adapted with permission from \citet{2008Sci...319..927G}, Figure 1}].}
\label{fig:lightcurves}
\end{figure}

The planet produced an anomaly in the event light curve that lasted about a day, but the signal was picked up and confirmed by four different telescopes independently (see Figure \ref{fig:lightcurves}a). The signal itself was generated by the source passing over a wide planetary caustic, which caused a brief boost in the observed magnification about 10 days after the main peak of the event.

\subsubsection{A Jupiter-Saturn Analog}
In 2008, \citet{2008Sci...319..927G} published the first microlensing discovery of a two-planet system (\mbox{see Figure \ref{fig:lightcurves}b}). The host is about half as massive as the Sun and the two gas giant planets are $\sim$0.71 and $\sim$0.27 times the mass of Jupiter with orbital separations of $\sim$2.3 and $\sim$4.6 AU respectively, both beyond the star's snow line. The system therefore appears like a scaled-down version of our Solar system with regard to two of the outer giant planets. Even though only two planets were detected, the~presence of more planets with very short (<0.4 AU) or very long (>10 AU) orbits cannot be excluded. 

The estimated mass of the host was confirmed in 2010 by \citet{2010ApJ...713..837B}, who used the adaptive optics system of the Keck 10 m telescope to measure the flux of the lens in the $H$-band.

\subsubsection{Limits on the Frequency of Planets}
Each data point in a given light curve corresponds to a particular position of the source. As the event progresses, the images of the source generated by the lensing effect scan the periphery of the lens, their size growing larger the shorter the apparent lens-source separation (see Figure \ref{fig:single_lens}). Since~the observed magnification is given by dividing the surface area of the images by the surface area of the unlensed source, any perturbations to one of the images caused by the presence of a planet at the corresponding location will affect the measured magnification and produce a deviant data point. To~state it differently, each data point that does not deviate from the general shape of the light curve of a single lens, excludes the possibility that there is a planet at the location of the images at the time of observation. This means that for events with no observed planetary anomalies, it is possible to place limits on the presence of planets in the system.

The limits obtained from analyzing events with no observed deviations can be compared with the actual detections to place limits on the number and frequency of planets in the Galaxy \citep{2000ApJ...535..176A,2010ApJ...710.1641S,2011MNRAS.411....2D,2018AcA....68....1U}. Early~attempts to do this, even before any microlensing planets were reported, concluded that less than 1/3 of all stars host Jupiter-mass planets between $\sim$1--4 AU \citep{2002ApJ...566..463G,2004MNRAS.351..967S}. The light curves available for those studies had a mean sampling frequency of $\sim$1 day, so it was not possible to derive meaningful limits for lower-mass planets.

In a 2010 article, \citet{2010ApJ...720.1073G} derived a planet frequency beyond the snow line from microlensing that was a factor 7 larger than the one derived from radial velocity studies for planets with much shorter periods, from 2 to 2000 days. This difference was consistent with the gradient derived from radial velocity results when these were extrapolated well beyond the separations they measured. In the same study, Gould also concluded that only about 1/6 of all planet-hosting stars host planets analogous to our Solar system.

In an independent analysis published in 2012, \citet{2012Natur.481..167C} arrived at an estimate for the fraction of bound planets at distances of 0.5--10 AU from their host stars. He found that only $17^{+6}_{-9}$\% of stars host Jupiter-mass planets (in the mass range 0.3--10 $M_{\rm Jupiter}$), but also that Neptunes (10--30 $M_{\rm Earth}$) and super-Earths (5--10 $M_{\rm Earth}$) are much more common beyond the snow line; their respective abundances being $52^{+22}_{-29}$\% and $62^{+35}_{-37}$\%. The remarkable conclusion was that, on average, every star in the Galaxy hosts a planet, and that planets are the rule rather than the exception. Although the analysis has received some criticism for relying on a small sample of events and for its use of Bayesian methods to convert from mass ratios to masses \citep{2016ApJ...833..145S}, the main results have been confirmed by more recent studies \citep{2016MNRAS.457.4089S,2016MNRAS.457.1320T}. Interestingly, a 2016 analysis by \citet{2016ApJ...833..145S} which relied exclusively on MOA-II microlensing survey data obtained between 2007 and 2012, reported consistent results but found a broken power-law mass function with a change of slope at $q \sim 10^{-4}$. This ``turnover'' in the mass function was recently confirmed by \citet{2018AcA....68....1U}. Concrete numbers for the required sample sizes for constraining planetary mass functions are given in \citet{2011MNRAS.411....2D}.

\section{Strengths and Weaknesses of the Microlensing Method}
Having reviewed the particulars of the method, we can summarize its strengths and address its~weaknesses.

\subsection{Weaknesses}
\begin{itemize}[leftmargin=*,labelsep=5.5mm]
\item{The probability of observing planet-lensing events is low, $\sim$$10^{-8}$ per star. In practice, this does not constitute a problem because surveys monitor about a billion stars regularly and there are $\sim$3--5 new microlensing planets discovered every year.}
\item{Planetary anomalies last only a few hours for Earths and a few days for Jupiters. Dozens of telescopes around the world coordinate their efforts to provide high-cadence observations of even the briefest anomalies.}
\item{Microlensing planets are too far away and are not good targets for searching for life with next-generation space missions. Although these planets cannot be the studied further, they~provide the missing piece of the puzzle when it comes to understanding how planetary systems form and evolve: what happens beyond the snow line.}
\item{Microlensing events are one-off occurrences, with no possibility of observing them again. Yet their light curves are observed from dozens of telescopes around the world, providing independent confirmation of the features detected. They are typically very well sampled, so anomalies can be well constrained.}
\item{It cannot find planets very close ($\lesssim$0.4 AU) or too far ($\gtrsim$100 AU) from their host stars. Those~regimes are better explored by transits, radial velocity and direct imaging. Microlensing is uniquely capable of exploring intermediate distances.}
\end{itemize}
\subsection{Strengths}
\begin{itemize}[leftmargin=*,labelsep=5.5mm]
\item{It does not have a bias for nearby stars, thus it explores the true Galactic population of planets.}
\item{It does not have a bias for the type of host star and can equally well find planets around brown or red-dwarf stars, main-sequence stars, stellar remnants, or event detect planets that have been ejected from their systems and are no longer gravitationally bound to their stars.}
\item{It does not require many years of observations to discover a planet. The typical microlensing event lasts for less than a month and provides a `snapshot' of the system, with a good chance of revealing any planets close to the Einstein ring of the lens.}
\item{It is very cheap in terms of the resources it requires, as it doesn't rely on having access to massive telescopes. 1m-class telescopes are routinely used for the observations, and for the brighter events even smaller amateur telescopes have contributed useful data.}
\item{It is sensitive to Earth-mass planets from ground-based observations.}
\item{It is uniquely capable of finding planets at and beyond the snow line of their host stars.}
\end{itemize}

\section{Future Prospects}
The population of close-in planets with orbits of less than 1 AU has been extensively explored through ground and space-based transit surveys, such as the Kepler mission, and through the radial velocity method. This regime will continue to be investigated, with improved sensitivity to planets with lower masses, by the newly launched (18 April 2018) {\it Transiting Exoplanet Survey Satellite} (TESS) and the {\it PLAnetary Transits and Oscillations of stars} (PLATO) mission, set to launch in 2026. Meanwhile, ground-based direct imaging efforts using adaptive optics on large telescopes will keep on finding large planets at orbits of hundreds of AU. It then depends on microlensing observing campaigns to fill the gap at intermediate distances. Having a complete picture is crucial for testing models of planet formation and developing our understanding about how planetary systems originate and evolve.

While there are plans to continue microlensing observations using ground-based facilities in the near future, finding planets of Earth-mass and below, as well as increasing the overall rate of discoveries, can be better achieved by going to space.

The NASA WFIRST space mission, set to launch in the mid-2020s, includes a microlensing planet-finding program. Going to space for a microlensing survey comes with several advantages (besides a stable PSF and uninterrupted coverage). First, WFIRST will rival Kepler in  number of planets detected. Simulations suggest it will detect thousands of microlensing exoplanets, thereby significantly shortening the time required to understand the distribution of planets between $\sim$1--100 AU, even for low-mass planets \citep{2002ApJ...574..985B}. Second, provided the events are also observed from the ground, it is possible to routinely measure the parallax effect, placing tight constraints on the physical parameters of the event \citep{2012arXiv1208.4012G}. Third, it will be sensitive to, and is expected to discover, planets that have been ejected from their systems, commonly referred to as `rogue' or `solivagant' planets \citep{2017ApJ...841...86B}.

Throughout this review we considered the photometric signature of microlensing events, but it is worth mentioning that they can also be detected through their astrometric signature, i.e., through small shifts in the light centroid of the observed star as the event progresses \citep{1998ApJ...494L..23P,1999ApJ...522..512S,2000ApJ...534..213D,2001PASJ...53..233H}. At lens-source separations greater than $\sim \theta_E$, the centroid shift caused by astrometric microlensing is proportional to $1/u$ while the photometric magnification is proportional to $1/u^4$. Because the astrometric signature of a microlensing event precedes and lasts longer than the photometric one, it is possible to anticipate a photometric signal when an expected high magnification event has been detected astrometrically. Combining photometric and astrometric microlensing can be useful in resolving some types of binary-lens degeneracies, although this still remains to be demonstrated \citep{1999MNRAS.309..373H,1999ApJ...526..405H}.

\citet{1995AcA....45..345P} first pointed out in 1995 that it is possible to predict astrometric microlensing events from precise proper motions and positions of lens and source stars. In light of this, a number of independent studies have used existing catalogs to select faint high-proper motion stars and, by~comparing their trajectories with background stars, predict astrometric microlensing events (\mbox{e.g., \citet{2000ApJ...539..241S,2011A&A...536A..50P}}). By far the most accurate measurements of positions and proper motions to-date are those contained in the recent second data release of the Gaia space observatory (Gaia DR2). \citet{2018A&A...615L..11K}, {Kl{\"u}ter} {et~al.} \cite{2018arXiv180711077K} and \citet{2018arXiv180610003B} used them to predict astrometric events in the coming decades, until the end of the century. A small fraction of these are likely to produce detectable photometric deviations and will inevitably attract the interest of observers.

\textls[-15]{The prospects of resolving individual images of stellar microlensing events using long-baseline interferometric observations have been improving with every new generation of instruments. Such~observations could be used to estimate the angular Einstein radius $\theta_E$ directly from the images. Typical microlensing values ($M_L \sim$ 0.5--1$ M_{\oplus}$, $\pi_{\mathrm{rel}} \sim$ 0.03--0.5 mas) result in Einstein radii in the range $\theta_E \sim$ 0.3--2.0 mas; This is very close to the limits of current VLTI instrumentation, such as PIONIER and GRAVITY, which can already achieve resolutions between 2.5 and 3 mas at maximum baseline \citep{2001A&A...375..701D,2016MNRAS.458.2074C}.}

\textls[-15]{New ideas are also being considered to improve the accuracy of ground-based photometric microlensing observations using Lucky Imaging techniques that compensate for atmospheric turbulence and deliver significantly sharper images when compared to the performance of standard CCD cameras, but so far they have only been tested with a relatively narrow field of view \citep{2013MNRAS.432..702M,2016MNRAS.458.3248S}. \citet{2017arXiv170900244M} recently argued that by combining several such cameras, wide-field high-resolution imaging is possible, which would provide enough sensitivity to detect bodies down to Lunar-mass with ground-based microlensing observations. While the construction of such an instrument comes with many technical challenges, it would be able to detect microlensing events in the most crowded regions of the Galactic bulge and compete with and complement WFIRST observations for the faintest targets.}

%%%%%%%%%%%%%%%%%%%%%%%%%%%%%%%%%%%%%%%%%%

%%%%%%%%%%%%%%%%%%%%%%%%%%%%%%%%%%%%%%%%%%
\vspace{6pt} 

%%%%%%%%%%%%%%%%%%%%%%%%%%%%%%%%%%%%%%%%%%
%% optional
%\supplementary{The following are Available online at www.mdpi.com/link, Figure S1: title, Table S1: title, Video S1: title.}

\funding{The author acknowledges the support of DFG priority program SPP 1992 ``Exploring the Diversity of Extrasolar Planets'' (WA 1047/11-1).}

%%%%%%%%%%%%%%%%%%%%%%%%%%%%%%%%%%%%%%%%%%
\acknowledgments{I would like to thank Etienne Bachelet for preparing Figure \ref{fig:caustic_topologies}, Markus Hundertmark for providing feedback on the text and the three anonymous referees for making helpful comments, corrections and suggestions that have greatly improved this work.}

%%%%%%%%%%%%%%%%%%%%%%%%%%%%%%%%%%%%%%%%%%

%%%%%%%%%%%%%%%%%%%%%%%%%%%%%%%%%%%%%%%%%%
\conflictofinterests{The author declares no conflict of interest.} 

%%%%%%%%%%%%%%%%%%%%%%%%%%%%%%%%%%%%%%%%%%
%% optional
\abbreviations{The following abbreviations are used in this manuscript:\\

\noindent 
\begin{tabular}{@{}ll}
CCD &Charge-coupled device \\
 AU& Astronomical unit. The average distance between the Sun and the Earth\\
WFIRST & Wide Field Infrared Survey Telescope\\
 PSPL&Point source - Point lens \\
 MACHO&Massive compact halo object \\
 MOA&Microlensing observations in astrophysics \\
 OGLE& Optical gravitational lensing experiment\\
 KMTNet&Korea Microlensing Telescope Network \\
 CCD&Charge-Coupled Device \\
 EMCCD& Electron Multiplying Charge Coupled Device\\
VLTI & Very Large Telescope Interferometer\\
PIONIER &Precision Integrated-Optics Near-infrared Imaging ExpeRiment \\
 GRAVITY& a VLTI instrument for precision astrometry and interferometric imaging
\end{tabular}}

%$M_{\oplus}$: Mass of the Earth\\
%$M_{\odot}$: Mass of the Sun\\
%pc: Parsec. 3.26 light-years\\
%$\theta_E$: Angular Einstein radius of the lens\\
%$t_E$: Event timescale. The time it takes for the source to move by one Einstein radius\\
%$D_L$: Distance between observer and lens\\
%$D_S$: Distance between observer and source\\
%$D_{LS}$: Distance between lens and source

%%%%%%%%%%%%%%%%%%%%%%%%%%%%%%%%%%%%%%%%%%
%% optional
%\appendix

%\section{}
%All appendix sections must be cited in the main text. In the appendixes, Figures, Tables, etc. should be labeled starting with `A', e.g., Figure A1, Figure A2, etc. 

%%%%%%%%%%%%%%%%%%%%%%%%%%%%%%%%%%%%%%%%%%
\reftitle{References}

%%%%%%%%%%%%%%%%%%%%%%%%%%%%%%%%%%%%%%%%%%
%% optional
%\sampleavailability{Samples of the compounds ...... are available from the authors.}

%%%%%%%%%%%%%%%%%%%%%%%%%%%%%%%%%%%%%%%%%%
\end{document}